%% file: main.tex
\documentclass[10pt,letterpaper,conference,nofonttune]{IEEEtran}
\IEEEoverridecommandlockouts

\usepackage{cite}
\usepackage{amsmath,amssymb,amsfonts}
\usepackage{graphicx}
\usepackage{xspace}

\newcommand{\nostradamus}{NOSTRAdAMUS\xspace}

\usepackage[acronyms,nonumberlist,nopostdot,nomain,nogroupskip,acronymlists={hidden}]{glossaries}
\newglossary[algh]{hidden}{acrh}{acnh}{Hidden Acronyms}
\input{acronyms.tex}

\newif\ifarchwide
\archwidetrue

\begin{document}
\bstctlcite{IEEEexample:BSTcontrol}
\title{Improving 5G AI-RAN MCS Selection by\\Predicting Retransmissions}

\author{
Tamerlan Aghayev\IEEEauthorrefmark{1},
Maxime Elkael\IEEEauthorrefmark{1},
Michele Polese\IEEEauthorrefmark{1},
Reshma Prasad\IEEEauthorrefmark{1},
Salvatore D'Oro\IEEEauthorrefmark{1},\\
Yunseong Lee\IEEEauthorrefmark{2},
Koichiro Furueda\IEEEauthorrefmark{2},
Tommaso Melodia\IEEEauthorrefmark{1}\\
\IEEEauthorrefmark{1}Institute for Intelligent Networked Systems, Northeastern University, Boston, MA, U.S.A.\\
\{aghayev.t, m.elkael, m.polese, re.prasad, s.doro, t.melodia\}@northeastern.edu\\
\IEEEauthorrefmark{2}SoftBank Research Institute of Advanced Technology, Tokyo, Japan\\
\{yunseong.lee, koichiro.furueda\}@g.softbank.co.jp
}

\maketitle
\glsunset{nr}
\glsunset{gnb}
\glsunset{5g}
\glsunset{3gpp}

\begin{abstract}
\gls{la} in 5G NR is inherently reactive, relying on channel measurements and \gls{harq} feedback that may become quickly obsolete when the channel changes quickly. This data is also noisy, making it hard to track accurately, and has to be fed to real-time controllers with feedback-loop effects which are hard to troubleshoot. This explains why most practical deployments select simple but robust algorithms, which accept that the lag can leave the scheduler operating at overly aggressive or unnecessarily conservative rates, trading spectrum efficiency for predictable performance. In this paper, we improve on this status-quo with \nostradamus, a predictive \gls{la} framework which adds foresight to existing algorithms without replacing or redesigning them. \nostradamus predicts whether a retransmission will occur in the next radio frame from recent \gls{harq} history, and applies corrections to the \gls{mcs} selected by the underlying policy. We benchmark several \gls{ml} models and show that Gradient Boosting achieves 82.9\% accuracy overall with high-confidence interventions that are
correct 94.2\% of the time, and an inference latency of 5.5~$\mu$s. We train the model based on data collected \gls{ota} on the X5G testbed, using the open-source \gls{oai} 5G stack, NVIDIA Aerial, and COTS O-RAN \glspl{ru} and \glspl{ue}. The model is then deployed as a dApp, which we evaluate \gls{ota} as well as on various channels using the same testbed with hardware-in-the-loop channel emulators. This includes \gls{3gpp} TDL and CDL channels, single and multi antenna configurations, and pedestrian and vehicular mobility.
Our evaluation shows that without retraining, and across this variety of scenarios, the dApp augments two state-of-the-art \gls{la} algorithms, and increases goodput by up to 71.5\% while reducing retransmissions by up to 71.8\%. This demonstrates the robustness and generalization capabilities of our approach.
\end{abstract}

\glsresetall
\glsunset{nr}
\glsunset{gnb}
\glsunset{5g}
\glsunset{3gpp}
\glsunset{olla}

\section{Introduction}
\label{sec:introduction}
In \gls{5g} \gls{nr}, the scheduler selects a \gls{mcs} for every \gls{dl} \gls{tb}, aiming to optimize spectral efficiency against an inherently dynamic wireless channel. This selection is usually made by a two-loop \gls{la} process: an inner loop (ILLA) maps the \gls{cqi} of the \glspl{ue} to an initial \gls{mcs}; and an outer loop (OLLA) adjusts that mapping with a dynamic offset based on \gls{harq} feedback~\cite{pedersen2007frequency}. \gls{harq} provides acknowledgment (ACK/NACK) feedback on whether a transmitted \gls{tb} was successfully decoded, so that failed ones can be retransmitted.
Although this \gls{la} architecture is well-established in the industry, it is inherently reactive as actions are taken only after errors have occurred.

\begin{figure}[t]
\centering
\includegraphics[width=0.98\columnwidth]{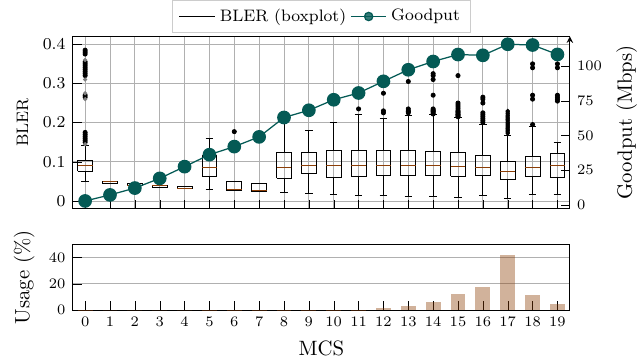}
\vspace{-.3cm}
\caption{Reactive \gls{mcs} selection for a static \gls{ue}. The loop selects an \gls{mcs} below 18 in 84\% of the time, while \gls{mcs}~18 is selected in less than 12\% of frames despite achieving a median \gls{bler} of 8.6\%, within the 10\% target.}
\vspace{-.3cm}
\label{fig:suboptimal_mcs}
\end{figure}

Unfortunately, being reactive is known to result in suboptimal \gls{mcs} selection when the channel varies fast~\cite{wang2026lolla, ye2023drlla}.
For example, when the most recent \gls{cqi} (reported periodically) indicates good channel conditions, ILLA maps to a high \gls{mcs}. However, if the channel degrades suddenly due to fast fading or blockage, a high \gls{mcs} might lead to decoding errors. In the meantime, the outer loop applies an adjustment derived from earlier positive \gls{harq} feedback and therefore cannot immediately account for this change. This is especially a problem in \gls{tdd} deployments, typical in 5G, as \gls{harq} and \gls{cqi} feedback are delayed, since the \gls{ue} needs to wait for \gls{ul} slots to report about the reception in \gls{dl}. As a result, the selected \gls{mcs} can be too aggressive for the current channel, causing failed transmissions and \gls{harq} retransmissions. The same also applies when the channel improves but the reactive loop remains at a conservative \gls{mcs} because its adjustment still reflects earlier decoding failures, leaving available link capacity unused. Figure~\ref{fig:suboptimal_mcs} shows that in our \gls{ota} indoor deployment, this problem occurs even for a static \gls{ue} in \gls{los} with the \gls{ru}. We use the industry-standard \gls{olla} algorithm~\cite{pedersen2007frequency}, parameterized to target 10\% of \gls{bler}. We observe that even with this easy scenario, the reactive loop spends 84\% of the time below the highest reliably supported operating point of \gls{mcs}~18, which achieves a median \gls{bler} of 8.6\%. This delay becomes particularly problematic in fast mobile scenarios~\cite{wang2026lolla}.

Recent work bridges this gap by redesigning \gls{la} algorithms, from enhanced \gls{olla} updates ~\cite{blanquez2016eolla,zhu2023nolla} to learned \gls{mcs} selection using bandits, supervised learning, and \gls{rl}~\cite{saxena2022rlla,tsipi2024mcs,ye2023drlla,you2026dcdqn}, as well as recent adaptive methods such as SALAD~\cite{wiesmayr2025salad} and policy-based LOLLA~\cite{wang2026lolla}. Another line of work predicts channel information used by \gls{la} through \gls{cqi} forecasting~\cite{yin2020cqi} or \gls{csi} prediction~\cite{diazruiz2025csi}.
Learned and enhanced-loop approaches replace or redesign the \gls{la} policy, while channel prediction only improves the information available to it. Neither provides a general way to add predictive foresight to an existing, well-understood loop without changing how it makes \gls{mcs} decisions. One must therefore either replace the loop or accept its lag.

We present \nostradamus, a predictive overlay framework that adds foresight to existing \gls{la} algorithms, thus improving their effectiveness without replacing them. Rather than learning a new \gls{mcs} policy from scratch, \nostradamus treats the existing \gls{la} algorithm as a black-box, and proactively corrects the \gls{mcs} it selects. Corrections are derived from \gls{ml} predictions of future retransmissions.
The \nostradamus overlay abstraction decouples the prediction from the \gls{la} design, allowing the same predictive mechanism to augment substantially different \gls{la} policies, from \gls{olla} to the \gls{sota} SALAD algorithm~\cite{wiesmayr2025salad}. To our knowledge, \nostradamus is the first predictive framework for \gls{la} designed as a plug-and-play overlay and demonstrated end-to-end with commercial radio hardware. 

\textbf{Main contributions.} We formulate the predictive \gls{la} overlaying problem as a binary classification over historical \gls{harq} data, enabling the scheduler to anticipate short-term decoding errors before reactive adaptation can respond. We then benchmark six \gls{ml} models and select Gradient Boosting~\cite{friedman2001greedy}, whose high-confidence interventions are correct $94.2\%$ of the time. We deploy \nostradamus as a dApp~\cite{lacava2025dapps} and validate it on the X5G testbed~\cite{villa2024x5g} with the open-source \gls{oai} stack, commercial \glspl{ru} and \glspl{ue}. We demonstrate \nostradamus by applying the same predictor to the OLLA and SALAD algorithms. We evaluate it across heterogeneous \gls{3gpp} \gls{tdl} and \gls{cdl} propagation models, \gls{siso} and \gls{mimo} configurations, pedestrian and vehicular mobility, and \gls{ota} experiments, achieving respectively up to $71.5\%$ and 60.9\% higher average goodput than OLLA and SALAD, while reducing retransmissions by up to $71.8\%$.

\newcommand{\archcaption}{\nostradamus architecture. The overlay dApp observes the host \gls{la} policy $\pi_0$, predicts first-retransmission activity for the next frame from the recent ratios $\rho$, and applies the confidence-gated adjustment $a_{u,n}$ of \eqref{eq:predictive_correction} to the selected \gls{mcs}.}

\ifarchwide
\begin{figure*}[t]
\centering
\includegraphics[width=\textwidth]{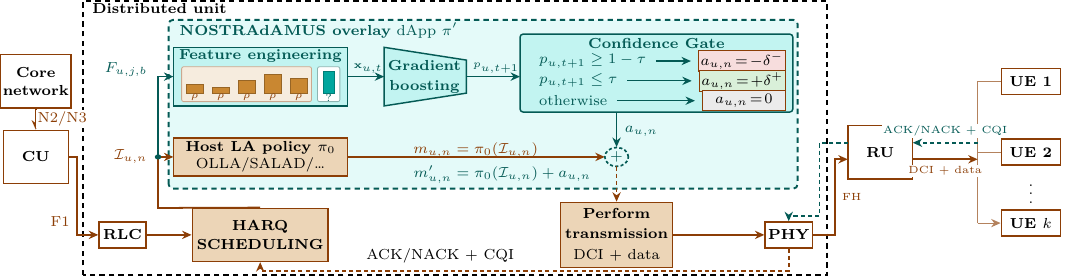}
\setlength{\abovecaptionskip}{-5pt}
\caption{\archcaption}
\label{fig:architecture}
\vspace{-.3cm}
\end{figure*}
\else
\begin{figure}[tb]
\centering
\includegraphics[width=0.98\columnwidth]{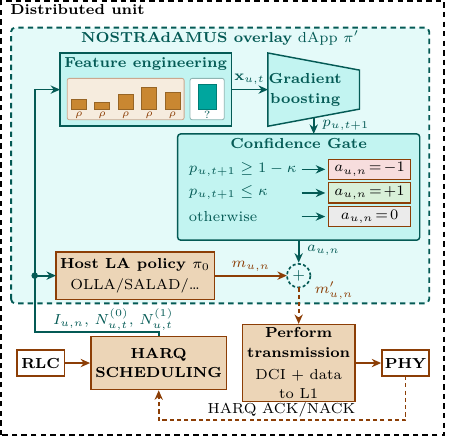}
\caption{\archcaption}
\setlength{\belowcaptionskip}{-5pt}
\setlength{\abovecaptionskip}{-5pt}
\label{fig:architecture}
\end{figure}
\fi

\section{System Model}
\label{sec:sysmodel}

We consider the downlink of a \gls{5g} \gls{gnb} serving a set of \glspl{ue} $\mathcal{U}=\{1,\ldots,N_{\mathrm{UE}}\}$. Time is divided into frames indexed by $t\in\mathbb{Z}_{\geq 0}$, each containing a set $\mathcal{D}_t$ of downlink slot indices as per the configured \gls{tdd} pattern (e.g., 6DS3U). We index downlink slots so that each slot $n\in\mathcal{D}_t$ belongs to a unique frame $t$. At each downlink slot $n$, the \gls{gnb} selects a set $\mathcal{U}_n\subseteq\mathcal{U}$ of \glspl{ue} to serve, allocates radio resources, schedules pending \gls{harq} retransmissions, and determines an \gls{mcs} for each \gls{ue} scheduled for a new \gls{tb}.
We consider \gls{harq} priority scheduling, where pending \gls{harq} retransmissions are scheduled before new-data transmissions.

We start by formalizing the link adaptation problem, in which the \gls{gnb} needs to select the \gls{mcs} for each scheduled \gls{ue}.
For \gls{ue} {$u\in\mathcal{U}_n$ scheduled in downlink slot $n\in\mathcal{D}_t$}, let $\mathcal{I}_{u,n}\in\mathcal{I}$ denote the information available to an \gls{la} policy when making its decision for \gls{ue} $u$ in slot $n$. $\mathcal{I}$ is the space of possible information, including \glspl{kpi} describing the recent conditions (e.g., throughput, \gls{cqi}, \gls{snr}, \gls{bler}, previous \gls{mcs}) and any internal state maintained by the policy (e.g., an \gls{olla} offset). An \gls{la} policy $\pi$ is a mapping $\pi:\mathcal{I}\longrightarrow\mathcal{M}$ with $\mathcal{M}=\{m_{\min},m_{\min}+1,\ldots,m_{\max}\}$
such that the \gls{mcs} selected for \gls{ue} $u$ in slot $n$ is $m_{u,n}=\pi\!\left(\mathcal{I}_{u,n}\right)\in\mathcal{M}$.

This \gls{mcs} decision determines the rate-reliability tradeoff of the transmission. For a given resource allocation, a higher \gls{mcs} carries a larger \gls{tb}, but also increases the probability of retransmissions. A lower \gls{mcs} reduces this risk at the cost of carrying fewer information bits over the same resources. 

Consider the set $\mathcal{B}_{u,t}$ of new \glspl{tb} whose initial transmissions are scheduled to \gls{ue} $u$ during frame $t$. As with downlink slots, we index \glspl{tb} so that each $i\in\mathcal{B}_{u,t}$ belongs to a unique frame $t$, hence the pair $(u,i)$ identifies a \gls{tb} uniquely. For each $i\in\mathcal{B}_{u,t}$, let $B_{u,i}>0$ denote the payload size of the \gls{tb}, fixed at its initial transmission and unchanged throughout its \gls{harq} chain. 

Let $R$ be the maximum number of \gls{harq} retransmissions configured for the
cell, and let $A_{u,i}\in\{1,\ldots,R+1\}$ denote the realized number of transmission attempts for \gls{tb} $i$, including its initial transmission. We index these attempts by $r=0,\ldots,A_{u,i}-1$, where $r=0$ denotes the initial transmission and $r\geq1$ a retransmission. After the \gls{harq} chain terminates, let $S_{u,i}\in\{0,1\}$ indicate whether the \gls{tb} was successfully decoded, and let $C^{(r)}_{u,i}>0$ denote the number of \glspl{prb} allocated to attempt $r$.

To capture this rate-reliability tradeoff over time, let $\mathcal{B}_T = \bigcup_{u\in\mathcal{U}}\bigcup_{t<T}\left\{(u,i) : i \in \mathcal{B}_{u,t}\right\}$ collect the \glspl{tb} initiated during the first $T$ frames. We characterize the resource efficiency of an \gls{la} policy by the long-run ratio of successfully delivered information bits to the radio resources consumed across all transmission attempts:

\begin{equation}
    J(\pi) = \liminf_{T \to \infty}\;
    \frac{\mathbb{E}_\pi\!\left\{\sum_{(u,i) \in \mathcal{B}_T} B_{u,i} S_{u,i}\right\}}
        {\mathbb{E}_\pi\!\left\{\sum_{(u,i) \in \mathcal{B}_T}\; \sum_{r=0}^{A_{u,i}-1} C^{(r)}_{u,i}\right\}},
\label{eq:objective}
\end{equation}
\noindent
where expectation is taken over stochastic elements (e.g., channel, traffic, decoding outcomes) under policy $\pi$. Thus, $J(\pi)$ measures the long-run useful information delivered per allocated \gls{prb}.

Ideally, $\pi$ would track the instantaneous channel quality to continuously operate at the Pareto frontier of this trade-off. This, however, requires the ability to track changes of $\mathcal{I}_{u,n}$ instantaneously despite its noisy nature. Typical algorithms struggle with this, which we address in the following section.

\section{\nostradamus}
\label{sec:nostradamus}

Figure~\ref{fig:architecture} shows the \nostradamus architecture. Starting from a reference \gls{la} policy $\pi_0:\mathcal{I}\rightarrow\mathcal{M}$, \nostradamus  constructs a new policy $\pi'$ that adjusts the \gls{mcs} decision of $\pi_0$ using \gls{ml}-based predictions derived from recent \gls{harq} activity. The adjusted \gls{mcs} decision is

\begin{equation} 
m'_{u,n}= \pi'\!\left(\pi_0(\mathcal{I}_{u,n})\right) = \pi_0(\mathcal{I}_{u,n}) + a_{u,n} \in \mathcal{M}.
\label{eq:nostradamus_policy} 
\end{equation}
\noindent
where $a_{u,n}$ is the \gls{mcs} adjustment described below. 

Consider a \gls{tb} that successfully decodes after $A_{u,i}$ attempts. \nostradamus only considers the outcome of the first transmission for learning purposes (i.e., it discards the outcomes of transmissions 2 to $A_{u,i}$), as successive attempts do not capture the channel state as accurately, since they also factor in \gls{harq} chase combining, in which \glspl{ue} use previously failed transmissions to increase their coding gain.

Let $F_{u,j,b}\in\{0,1\}$ denote the \gls{harq} outcome of the first transmission of the $b$-th new \gls{tb} in frame $j$, with 0 and 1 respectively meaning ACK and NACK.

For a trailing window of $W\geq1$ frames, in which the count of new downlink transmissions for \gls{ue} $u$ is $d_{u,1},\ldots,d_{u,W}$ the first-transmission \gls{bler} is:

\begin{equation}
\rho^{(W)}_{u,t} = \Big(1 \big/ \sum_{j=1}^{W} d_{u,j}\Big)
\times \sum_{j=1}^{W} \sum_{b=1}^{d_{u,j}} F_{u,j,b}.
\label{eq:retx_ratio}
\end{equation}
\noindent

Note that windows without initial transmissions ($d_{u,j}=0$ for every $j$ in the window) are treated as missing observations. In the remainder of this paper, \gls{bler} refers to the first transmission \gls{bler}.

\nostradamus predicts whether \gls{ue} $u$'s new \glspl{tb} will likely require retransmission in
the next frame. For a frame satisfying $d_{u,t+1}>0$, the label is $y_{u,t+1} = \mathbf{1}\!\left\{ \rho^{(1)}_{u,t+1}>0 \right\}$.

At the end of frame $t$, let $\mathbf{x}_{u,t}\in\mathbb{R}^{H}$ collect the $H$ most recent values of $\rho^{(W)}_{u,k}$, $k\leq t$,
and let $f_{\phi}:\mathbb{R}^{H}\rightarrow[0,1]$ denote the learned predictor (see Sec.~\ref{sec:training_deployment}). Its output is

\begin{equation}
p_{u,t+1} = f_{\phi}\!\left(\mathbf{x}_{u,t}\right)
\approx \Pr\!\left(y_{u,t+1}=1 \mid \mathbf{x}_{u,t}\right).
\label{eq:prediction}
\end{equation}

Inference runs once per \gls{ue} per frame, at the end of frame $t$, and $p_{u,t+1}$ is then held for every \gls{mcs} decision in frame $t+1$. Each prediction is performed independently for each \gls{ue} using that \gls{ue}'s own retransmission history. If fewer than $H$ valid observations are available, or inference misses its deadline, \nostradamus abstains ($a_{u,n}=0$) and $m'_{u,n}=\pi_0(\mathcal{I}_{u,n})$.

Given the predicted probability $p_{u,t+1}$ in \eqref{eq:prediction}, the adjustment applied to a \gls{mcs} in slot $n\in\mathcal{D}_{t+1}$ is

\begin{equation}
a_{u,n} =
\begin{cases}
-\delta^{-}, & p_{u,t+1} \geq 1-\tau,\quad
\pi_0(\mathcal{I}_{u,n}) - \delta^{-} \geq m_{\min},\\
+\delta^{+}, & p_{u,t+1} \leq \tau,\quad
\pi_0(\mathcal{I}_{u,n}) + \delta^{+} \leq m_{\max},\\
0, & \text{otherwise}.
\end{cases}
\label{eq:predictive_correction}
\end{equation}
where $\tau\in\left[0,\tfrac{1}{2}\right)$ is a confidence threshold and $\delta^{-}, \delta^{+}\in\mathbb{Z}_{>0}$ are configurable downward/upward \gls{mcs} corrections.

\section{Training and Deployment}
\label{sec:training_deployment}

To train \nostradamus we use real world data collected on X5G~\cite{villa2024x5g} using the AutoRAN~\cite{maxenti2025autoranautomatedzerotouchopen} automation framework. The \gls{ran} runs on a Gigabyte E251 server equipped with an Intel Xeon 6240R \gls{cpu} and an NVIDIA A100 GPU. We use a Foxconn RPQN configured for 2$\times$2 \gls{mimo} operation over a 40~MHz bandwidth, with a 6DS3U \gls{tdd} pattern and numerology 1. The core network is Open5GS. An \gls{oai} \gls{gnb} with the NVIDIA Aerial PHY layer serves saturated \gls{dl} \glsunset{udp}\gls{udp} iPerf traffic to a Samsung S25 \gls{ue}, carried by a user walking indoors (inside our lab).
The walk exercises a wide channel condition range: \gls{rsrp} spans -107 to -53 dBm between its 5th and 95th percentiles and the scheduler visits every \gls{mcs} from 0 to 27. Resulting 2.4 million slot-level samples are aggregated to frame granularity, yielding 172,324 frame observations. The dataset is chronologically split into $70\%$ training, $15\%$ validation, and $15\%$ testing. The held-out test set contains $25{,}805$ samples,
of which $39.3\%$ are positive.

\begin{figure}[!t]
\centering
\includegraphics[width=0.98\columnwidth]{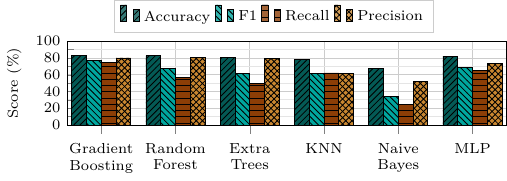}
\vspace{-.3cm}
\caption{\gls{ml} model benchmarking.}
\label{fig:ml_comparison_results}
\vspace{-.3cm}
\end{figure}

We benchmark candidate models spanning probabilistic (Naive Bayes), distance-based (k-Nearest Neighbors), tree-ensemble (Random Forest, Extra Trees, Gradient Boosting), and neural (Multi-Layer Perceptron) families on the same feature vector. Figure~\ref{fig:ml_comparison_results} summarizes accuracy, precision, recall, and F1 across models. Tree ensembles dominate the benchmark, with Gradient Boosting best overall.

\begin{figure}[!t]
\centering
\includegraphics[width=0.98\columnwidth]{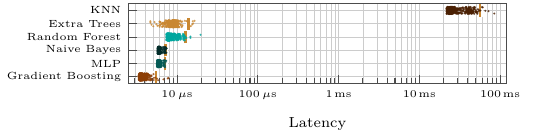}
\vspace{-.3cm}
\caption{Per-UE inference latency of the six \gls{ml} models in \gls{onnx}, 10{,}000 runs, amber bar marks the 99th percentile.
}
\label{fig:inference_latency}
\vspace{-.45cm}
\end{figure}

Because \nostradamus operates in the scheduling loop as a dApp, inference must complete in time as discussed in Sec.~\ref{sec:nostradamus}. We export the trained models to \glsunset{onnx}\gls{onnx} to support low-overhead real-time inference and evaluate their $99$th-percentile latency over $10{,}000$ runs (Fig.~\ref{fig:inference_latency}).
Gradient Boosting achieves the lowest inference latency at $5.5~\mu$s and, together with its predictive performance, is therefore selected for deployment.

\begin{figure}[!b]
\centering
\vspace{-.45cm}
\includegraphics{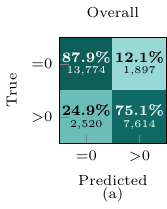}\hspace{5pt}%
\includegraphics{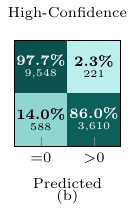}\hspace{7pt}%
\includegraphics{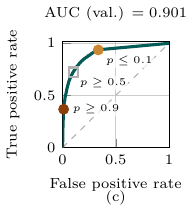}
\setlength{\belowcaptionskip}{-5pt}
\setlength{\abovecaptionskip}{-5pt}
\caption{Confusion matrices, (a) full test set and (b) high-confidence subset, (c) \gls{roc} on the validation set.}
\label{fig:gb_confusion_matrices}
\end{figure}
Figure~\ref{fig:gb_confusion_matrices} summarizes the selected Gradient Boosting model, which achieves $82.9\%$ accuracy and $75.1\%$ recall on the full test set. For all experiments, we set $W=10$, $H=5$, $\delta^{-}$= $\delta^{+}$= 1, and select $\tau=0.1$ on the validation set as a balance between confidence and intervention coverage (motivated by Fig.~\ref{fig:gb_confusion_matrices}(c)). On the test set, our confidence gate intervenes on $54.1\%$ of samples, achieving $94.2\%$ precision.

\begin{figure}[!t]
\centering
\includegraphics[width=0.88\columnwidth]{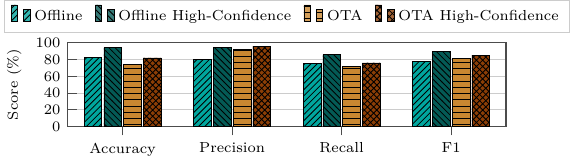}
\vspace{-.3cm}
\caption{Offline vs.\ deployment metrics, overall and high-confidence.}
\vspace{-.3cm}
\label{fig:training_vs_deployment}
\end{figure}

\begin{figure*}[t]
\centering
\includegraphics[width=\textwidth]{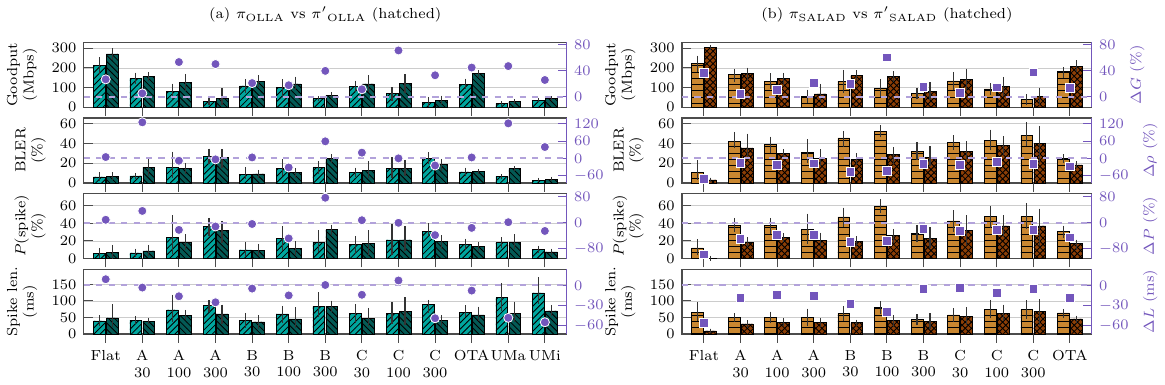}
\setlength{\belowcaptionskip}{-5pt}
\setlength{\abovecaptionskip}{-15pt}
\caption{Summary across all scenarios for base policy $\pi_0$ and \nostradamus overlay $\pi'$, with change $\Delta$ (right axes) and 95\% bootstrap intervals.}
\vspace{-.3cm}
\label{fig:result_summary}
\end{figure*}

Figure~\ref{fig:training_vs_deployment} compares prediction accuracy for offline and \gls{ota} regimes. Overall accuracy drops
by $8.9$ and by $12.2$ points on the high-confidence subset, while precision increases in both cases ($+11.9$ and $+1.8$ points) as recall decreases, and higher precision reduces conservative steps on clean frames. 

\section{Experimental Evaluation}
\label{sec:results}

In this section, we evaluate whether (i) the learned overlay generalizes beyond the channel conditions seen during training;
(ii) it adapts to user mobility; and (iii)
we validate the complete closed-loop system \gls{ota} on the X5G testbed with commercial radio hardware. In all three cases, we use the same Gradient Boosting model trained on the \gls{ota} traces of Sec.~\ref{sec:training_deployment}, without retraining.
Figure~\ref{fig:result_summary} summarizes goodput, \gls{bler}, and \gls{bler}-spike (we count a spike for each frame where \gls{bler} exceeds $0.5$) behavior across all scenarios.

\textbf{OpenAirLink: TDL-A/B/C.}
For controlled, repeatable single-antenna fading tests we use OpenAirLink (OAL), an open \gls{usrp}-based programmable channel emulator~\cite{deshpande2024openairlink}.
An \gls{oai} \gls{gnb} drives a Foxconn RPQN \gls{ru} inside a shielded enclosure, with the RF path between the \gls{ru} and a Sierra Wireless EM9293 \gls{cots} \gls{ue} passing through OAL running on a \gls{usrp} X410. OAL applies controlled attenuation and \gls{3gpp} fading profiles to the bidirectional radio link.

\begin{figure}[!ht]
\centering
\includegraphics[width=0.98\columnwidth]{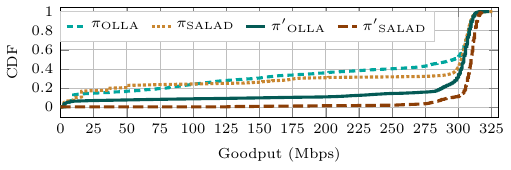}
\vspace{-.3cm}
\caption{Goodput \gls{cdf} on the single-tap flat channel.}
\label{fig:lchem_clean}
\vspace{-.3cm}
\end{figure}

To show \nostradamus's benefits even in static conditions, we start our evaluation with a single-tap flat channel. Figure~\ref{fig:lchem_clean} shows that
\nostradamus improves mean goodput
by $27.1\%$ over \gls{olla} at a nearly unchanged \gls{bler} ($+4.6\%$) and
by $37.2\%$ over SALAD while decreasing the \gls{bler} by $71.8\%$ (Fig.~\ref{fig:result_summary}). This means that even if the \gls{mcs} that maximizes goodput should be fixed, iterative processes like \gls{olla} and SALAD cannot converge to it stably.

\begin{figure}[!ht]
\centering
\includegraphics[width=0.98\columnwidth]{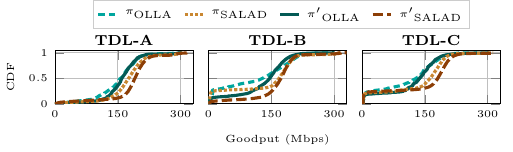}
\vspace{-.3cm}
\caption{Goodput \glspl{cdf} at a fixed 30~ns delay spread.}%
\vspace{-.5cm}
\label{fig:trio_channel}
\end{figure}

To understand if these gains are generalizable, we deploy \nostradamus, without retraining, on \gls{3gpp} TDL-A, TDL-B, and TDL-C channel models~\cite{3gpp_901} at a fixed delay spread of 30~ns. 
Although \nostradamus operates in completely unseen channel conditions with different delay profiles, Fig.~\ref{fig:trio_channel} shows mean goodput improvements in all three scenarios: by $5.5$--$21.3\%$ over \gls{olla} and $5.1$--$20.4\%$ over SALAD.
Thus, \nostradamus generalizes and improves both hosts on three unseen delay profiles, with no per-channel tuning.

\begin{figure}[!b]
\centering
\includegraphics[width=0.98\columnwidth]{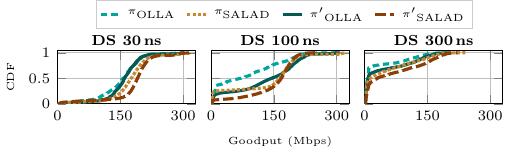}
\vspace{-.3cm}
\caption{Goodput \glspl{cdf} for TDL-A as the delay spread grows.}
\vspace{-.3cm}
\label{fig:trio_ds}
\end{figure}

We continue our cross-channel generalization study and evaluate what happens when the channel degrades the feedback the base policy depends on (e.g., \gls{cqi}) by fixing TDL-A and sweeping the delay spread from 30 to 300~ns (Fig.~\ref{fig:trio_ds}). A longer spread narrows the coherence bandwidth, so a single wideband \gls{cqi} becomes less representative of the increasingly uneven per-\gls{prb} channel. Both base policies collapse: mean goodput falls by $78.7\%$
for \gls{olla} and
$65.8\%$ for SALAD between DS 30 and DS 300. \nostradamus does not remove this loss, but its goodput gains persist, from $5.5\%$ to $53.7\%$ over \gls{olla} and from $5.1\%$ to $21.6\%$ over SALAD. As we can observe in Fig.~\ref{fig:trio_ds}, the base policies concentrate more on lower \gls{mcs} values as the delay spread grows, while the horizontal gap between each base and its overlay does not shrink. Thus, predictions of \nostradamus remain useful when frequency selectivity degrades the feedback \gls{la} depends on. 

We also repeat the delay spread sweep for TDL-B and TDL-C. Fig.~\ref{fig:result_summary} shows that \nostradamus generalizes across different channel profiles and improves goodput in all ten scenarios. Over \gls{olla}, the median gain is $30.2\%$ (range $5.5$--$71.5\%$) and $17.9\%$ (range $5.1$--$60.9\%$) over SALAD.

\begin{figure}[!t]
\centering
\includegraphics[width=0.98\columnwidth]{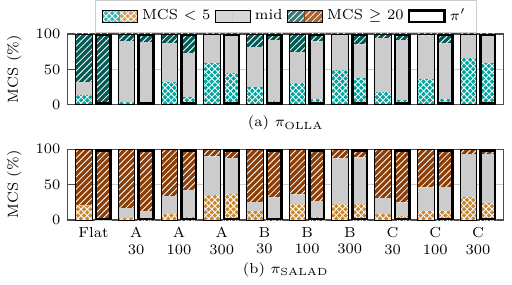}
\vspace{-.3cm}
\caption{First-attempt \gls{mcs} distribution over all OAL experiments, base solid and \nostradamus hatched: (a) on the \gls{olla} base, (b) on the SALAD base.}
\vspace{-.3cm}
\label{fig:mcs_distribution}
\end{figure}

\begin{figure}[!t]
\centering
\includegraphics[width=0.98\columnwidth]{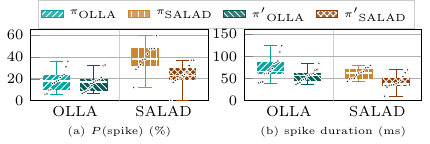}
\vspace{-.3cm}
\caption{\gls{bler} spikes' probability (a) and its mean duration (b).}
\vspace{-.5cm}
\label{fig:bursts}
\end{figure}

Overall, \nostradamus benefits the two base policies differently. With \gls{olla}, the \gls{bler} remains unchanged ($-1.2$\%), while the first-attempt transmissions below \gls{mcs}~5 fall by $73.4\%$
(Fig.~\ref{fig:mcs_distribution}(a)). Thus, the overlay prevents delayed feedback from keeping \gls{olla} at overly conservative rates. SALAD already operates at higher \gls{mcs} values, and the overlay changes its high-\gls{mcs} share by only $3.1$\% at the median. Instead, it reduces the \gls{bler} by $12.0$--$71.8$\% in every scenario and lowers the \gls{bler} spike probability
by $41.7\%$ and its duration
by $38.4\%$ (Fig.~\ref{fig:bursts}). In short, \nostradamus accelerates rate recovery for \gls{olla}, and reduces retransmissions for SALAD.

\textbf{PROPSIM: CDL-C UMi/UMa.} We now add mobility with a \gls{mimo} evaluation under the \gls{3gpp}~38.901 CDL-C model on the Keysight PROPSIM FS64. We use UMi at 1.5~m/s (pedestrian) and UMa at 30~km/h (vehicular), directly stressing the failure mode introduced in Sec.~\ref{sec:introduction}, i.e., the channel varies while a reactive policy waits for delayed \gls{harq} outcomes.
In both scenarios, \nostradamus holds a consistent advantage. 

\begin{figure}[!t]
\centering
\includegraphics[width=0.94\columnwidth]{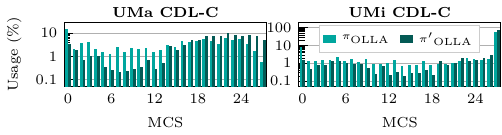}
\vspace{-.3cm}
\caption{\gls{mcs} usage under CDL-C mobility. Log scaled y-axis.}
\vspace{-.3cm}
\label{fig:propsim_mcs}
\end{figure}

In \gls{olla}, deep fades cause \gls{mcs} to drop fast with a slow recovery rate. This causes \gls{mcs}~0 to be selected $15.2\%$ and $8.2\%$ of the time in UMa and UMi, respectively.
\nostradamus avoids \gls{mcs}~0 almost entirely, shrinking the sub-\gls{mcs}-5 region by
$82.9$\% on UMa, and by $75.5$\% on UMi, and raising the \gls{mcs}~$\geq20$ share
by $108.3$\% and $27.0$\% (Fig.~\ref{fig:propsim_mcs}). 

\begin{figure}[!ht]
\centering
\includegraphics[width=0.95\columnwidth]{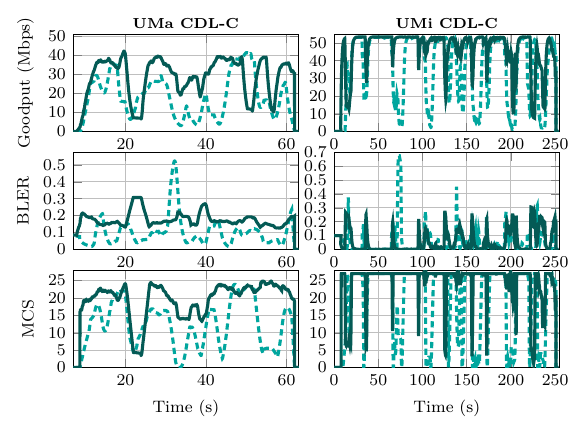}
\vspace{-.3cm}
\caption{Time series under CDL-C mobility. OLLA (dashed), overlay (solid).}
\vspace{-.3cm}
\label{fig:propsim_ts}
\end{figure}

The time series in Fig.~\ref{fig:propsim_ts} better explain this gain.
\nostradamus restores the high-rate regime faster and holds it for longer continuous intervals, whereas \gls{olla} over-corrects into prolonged low-\gls{mcs} episodes.
In short, our CDL-C experiments show that under mobility, \nostradamus contributes by reducing the recovery time, and its value grows with the rate at which the channel forces recoveries.

\begin{figure}[!t]
\centering
\includegraphics[width=0.98\columnwidth]{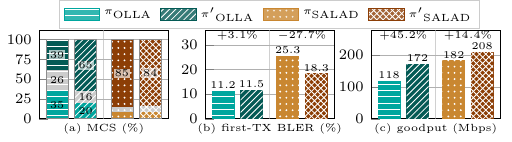}
\vspace{-.3cm}
\caption{\gls{ota} results: (a) first-attempt \gls{mcs}, (b) BLER, (c) goodput.}
\vspace{-.5cm}
\label{fig:ota_mcs}
\end{figure}

\textbf{X5G: Over the Air.} The \gls{ota} evaluation uses the same testbed and configuration as the training campaign (Sec.~\ref{sec:training_deployment}), with a bandwidth of 100~MHz.
Overall, our \gls{ota} experiments reproduce the trends observed under emulation. With \gls{olla}, the gain again comes from spending less time at lower \gls{mcs} regions, lifting goodput by $45.2\%$
with similar \gls{bler} ($+3.1\%$).
On SALAD the \gls{mcs} distribution is almost identical,
 with goodput gains ($+14.4\%$) resulting from reducing the \gls{bler} ($-27.7\%$) (Fig.~\ref{fig:ota_mcs}).%

\section{Conclusions \& Future Work}
\label{sec:conclusions}

We proposed \nostradamus, a novel plug-and-play predictive overlay that adds foresight to existing \gls{la} policies without replacing or redesigning them.

Using only short-term \gls{harq} history, a lightweight Gradient Boosting model predicts upcoming retransmission activity and proactively adjusts the \gls{mcs} selected by the underlying policy.
Across \gls{3gpp} \gls{tdl} and \gls{cdl} channels, SISO and MIMO configurations, pedestrian and vehicular mobility, and \gls{ota} deployment, the same predictor generalizes and improves both \gls{olla} and SALAD. Despite their different adaptation mechanisms, \nostradamus complements both policies by accelerating link recovery when \gls{olla} becomes overly conservative and reduces unnecessary retransmissions over SALAD. Overall, the overlay increases average goodput by up to $71.5\%$ and reduces retransmissions by up to $71.8\%$, demonstrating that predictive adaptation can be introduced as a lightweight layer on top of existing \gls{la} algorithms.

Future work will evaluate \nostradamus in larger multi-\gls{ue} deployments, extend the framework to uplink \gls{la}, and investigate its ability to augment additional base policies, including \gls{rl}-based approaches.

\bibliographystyle{IEEEtran}
\bibliography{references}

\end{document}

%% file: acronyms.tex
\newacronym{3gpp}{3GPP}{3rd Generation Partnership Project}
\newacronym{4g}{4G}{4th generation}
\newacronym{5g}{5G}{5th generation}
\newacronym{6g}{6G}{6th generation}
\newacronym{5gc}{5GC}{5G Core}
\newacronym{adc}{ADC}{Analog to Digital Converter}
\newacronym{aerpaw}{AERPAW}{Aerial Experimentation and Research Platform for Advanced Wireless}
\newacronym{ai}{AI}{Artificial Intelligence}
\newacronym{arq}{ARQ}{Automatic Repeat reQuest}
\newacronym{aimd}{AIMD}{Additive Increase Multiplicative Decrease}
\newacronym{am}{AM}{Acknowledged Mode}
\newacronym{amc}{AMC}{Adaptive Modulation and Coding}
\newacronym{amf}{AMF}{Access and Mobility Management Function}
\newacronym{aops}{AOPS}{Adaptive Order Prediction Scheduling}
\newacronym{api}{API}{Application Programming Interface}
\newacronym{apn}{APN}{Access Point Name}
\newacronym{ap}{AP}{Application Protocol}
\newacronym{aqm}{AQM}{Active Queue Management}
\newacronym{auc}{AUC}{Area Under the Curve}
\newacronym{ausf}{AUSF}{Authentication Server Function}
\newacronym{avc}{AVC}{Advanced Video Coding}
\newacronym{awgn}{AGWN}{Additive White Gaussian Noise}
\newacronym{balia}{BALIA}{Balanced Link Adaptation Algorithm}
\newacronym{bbu}{BBU}{Base Band Unit}
\newacronym{bdp}{BDP}{Bandwidth-Delay Product}
\newacronym{ber}{BER}{Bit Error Rate}
\newacronym{bf}{BF}{Beamforming}
\newacronym{bler}{BLER}{Block Error Rate}
\newacronym{brr}{BRR}{Bayesian Ridge Regressor}
\newacronym{bs}{BS}{Base Station}
\newacronym{bsr}{BSR}{Buffer Status Report}
\newacronym{bss}{BSS}{Business Support System}

\newacronym{ca}{CA}{Carrier Aggregation}
\newacronym{caas}{CaaS}{Connectivity-as-a-Service}
\newacronym{cb}{CB}{Code Block}
\newacronym{cc}{CC}{Congestion Control}
\newacronym{ccid}{CCID}{Congestion Control ID}
\newacronym{cco}{CC}{Carrier Component}
\newacronym{cdd}{CDD}{Cyclic Delay Diversity}
\newacronym{cdf}{CDF}{Cumulative Distribution Function}
\newacronym{cdn}{CDN}{Content Distribution Network}
\newacronym{cn}{CN}{Core Network}
\newacronym{codel}{CoDel}{Controlled Delay Management}
\newacronym{comac}{COMAC}{Converged Multi-Access and Core}
\newacronym{cord}{CORD}{Central Office Re-architected as a Datacenter}
\newacronym{cornet}{CORNET}{COgnitive Radio NETwork}
\newacronym{cosmos}{COSMOS}{Cloud Enhanced Open Software Defined Mobile Wireless Testbed for City-Scale Deployment}
\newacronym{cots}{COTS}{Commercial Off-the-Shelf}
\newacronym{cp}{CP}{Control Plane}
\newacronym{cyp}{CP}{Cyclic Prefix}
\newacronym{up}{UP}{User Plane}
\newacronym{cpu}{CPU}{Central Processing Unit}
\newacronym{cqi}{CQI}{Channel Quality Indicator}
\newacronym{cr}{CR}{Cognitive Radio}
\newacronym{cran}{CRAN}{Cloud \gls{ran}}
\newacronym{crs}{CRS}{Cell Reference Signal}
\newacronym{crc}{CRC}{Cyclic Redundancy Check}
\newacronym{csi}{CSI}{Channel State Information}
\newacronym{csirs}{CSI-RS}{Channel State Information - Reference Signal}
\newacronym{cu}{CU}{Central Unit}
\newacronym{d2tcp}{D$^2$TCP}{Deadline-aware Data center TCP}
\newacronym{d3}{D$^3$}{Deadline-Driven Delivery}
\newacronym{dac}{DAC}{Digital to Analog Converter}
\newacronym{dag}{DAG}{Directed Acyclic Graph}
\newacronym{das}{DAS}{Distributed Antenna System}
\newacronym{dash}{DASH}{Dynamic Adaptive Streaming over HTTP}
\newacronym{dc}{DC}{Dual Connectivity}
\newacronym{dccp}{DCCP}{Datagram Congestion Control Protocol}
\newacronym{dce}{DCE}{Direct Code Execution}
\newacronym{dci}{DCI}{Downlink Control Information}
\newacronym{dctcp}{DCTCP}{Data Center TCP}
\newacronym{dl}{DL}{Downlink}
\newacronym{dmr}{DMR}{Deadline Miss Ratio}
\newacronym{dmrs}{DMRS}{DeModulation Reference Signal}
\newacronym{drlcc}{DRL-CC}{Deep Reinforcement Learning Congestion Control}
\newacronym{drs}{DRS}{Discovery Reference Signal}
\newacronym{du}{DU}{Distributed Unit}
\newacronym{e2e}{E2E}{end-to-end}
\newacronym{ecaas}{ECaaS}{Edge-Cloud-as-a-Service}
\newacronym{ecn}{ECN}{Explicit Congestion Notification}
\newacronym{edf}{EDF}{Earliest Deadline First}
\newacronym{embb}{eMBB}{Enhanced Mobile Broadband}
\newacronym{empower}{EMPOWER}{EMpowering transatlantic PlatfOrms for advanced WirEless Research}
\newacronym{enb}{eNB}{evolved Node Base}
\newacronym{endc}{EN-DC}{E-UTRAN-\gls{nr} \gls{dc}}
\newacronym{epc}{EPC}{Evolved Packet Core}
\newacronym{eps}{EPS}{Evolved Packet System}
\newacronym{es}{ES}{Edge Server}
\newacronym{etsi}{ETSI}{European Telecommunications Standards Institute}
\newacronym[firstplural=Estimated Times of Arrival (ETAs)]{eta}{ETA}{Estimated Time of Arrival}
\newacronym{eutran}{E-UTRAN}{Evolved Universal Terrestrial Access Network}
\newacronym{faas}{FaaS}{Function-as-a-Service}
\newacronym{fapi}{FAPI}{Functional Application Platform Interface}
\newacronym{fdd}{FDD}{Frequency Division Duplexing}
\newacronym{fdm}{FDM}{Frequency Division Multiplexing}
\newacronym{fdma}{FDMA}{Frequency Division Multiple Access}
\newacronym{fed4fire}{FED4FIRE+}{Federation 4 Future Internet Research and Experimentation Plus}
\newacronym{fir}{FIR}{Finite Impulse Response}
\newacronym{fit}{FIT}{Future \acrlong{iot}}
\newacronym{fpga}{FPGA}{Field Programmable Gate Array}
\newacronym{fr2}{FR2}{Frequency Range 2}
\newacronym{fs}{FS}{Fast Switching}
\newacronym{fscc}{FSCC}{Flow Sharing Congestion Control}
\newacronym{ftp}{FTP}{File Transfer Protocol}
\newacronym{fw}{FW}{Flow Window}
\newacronym{ge}{GE}{Gaussian Elimination}
\newacronym{gnb}{gNB}{Next Generation Node Base}
\newacronym{gop}{GOP}{Group of Pictures}
\newacronym{gpr}{GPR}{Gaussian Process Regressor}
\newacronym{gpu}{GPU}{Graphics Processing Unit}
\newacronym{gtp}{GTP}{GPRS Tunneling Protocol}
\newacronym{gtpc}{GTP-C}{GPRS Tunnelling Protocol Control Plane}
\newacronym{gtpu}{GTP-U}{GPRS Tunnelling Protocol User Plane}
\newacronym{gtpv2c}{GTPv2-C}{\gls{gtp} v2 - Control}
\newacronym{gw}{GW}{Gateway}
\newacronym{harq}{HARQ}{Hybrid Automatic Repeat reQuest}
\newacronym{hetnet}{HetNet}{Heterogeneous Network}
\newacronym{hh}{HH}{Hard Handover}
\newacronym{hol}{HOL}{Head-of-Line}
\newacronym{hqf}{HQF}{Highest-quality-first}
\newacronym{hss}{HSS}{Home Subscription Server}
\newacronym{http}{HTTP}{HyperText Transfer Protocol}
\newacronym{ia}{IA}{Initial Access}
\newacronym{iab}{IAB}{Integrated Access and Backhaul}
\newacronym{ic}{IC}{Incident Command}
\newacronym{ietf}{IETF}{Internet Engineering Task Force}
\newacronym{ims}{IMS}{Infrastructure Management Service}
\newacronym{imsi}{IMSI}{International Mobile Subscriber Identity}
\newacronym{imt}{IMT}{International Mobile Telecommunication}
\newacronym{iot}{IoT}{Internet of Things}
\newacronym{ip}{IP}{Internet Protocol}
\newacronym{itu}{ITU}{International Telecommunication Union}
\newacronym{illa}{ILLA}{Inner Loop Link Adaptation}
\newacronym{ibler}{iBLER}{initial \gls{bler}}
\newacronym{kpi}{KPI}{Key Performance Indicator}
\newacronym{kpm}{KPM}{Key Performance Measurement}
\newacronym{kvm}{KVM}{Kernel-based Virtual Machine}
\newacronym{los}{LOS}{Line-of-Sight}
\newacronym{lsm}{LSM}{Link-to-System Mapping}
\newacronym{lstm}{LSTM}{Long Short Term Memory}
\newacronym{lte}{LTE}{Long Term Evolution}
\newacronym{lxc}{LXC}{Linux Container}
\newacronym{m2m}{M2M}{Machine to Machine}
\newacronym{mac}{MAC}{Medium Access Control}
\newacronym{manet}{MANET}{Mobile Ad Hoc Network}
\newacronym{mano}{MANO}{Management and Orchestration}
\newacronym{mc}{MC}{Multi-Connectivity}
\newacronym{mcc}{MCC}{Mobile Cloud Computing}
\newacronym{mchem}{MCHEM}{Massive Channel Emulator}
\newacronym{mcs}{MCS}{Modulation and Coding Scheme}
\newacronym{mec}{MEC}{Multi-access Edge Computing}
\newacronym{mec2}{MEC}{Mobile Edge Cloud}
\newacronym{mfc}{MFC}{Mobile Fog Computing}
\newacronym{mgen}{MGEN}{Multi-Generator}
\newacronym{mi}{MI}{Mutual Information}
\newacronym{mib}{MIB}{Master Information Block}
\newacronym{miesm}{MIESM}{Mutual Information Based Effective SINR}
\newacronym{mimo}{MIMO}{Multiple Input, Multiple Output}
\newacronym{ml}{ML}{Machine Learning}
\newacronym{mlr}{MLR}{Maximum-local-rate}
\newacronym[plural=\gls{mme}s,firstplural=Mobility Management Entities (MMEs)]{mme}{MME}{Mobility Management Entity}
\newacronym{mmtc}{mMTC}{Massive Machine-Type Communications}
\newacronym{mmwave}{mmWave}{millimeter wave}
\newacronym{mpdccp}{MP-DCCP}{Multipath Datagram Congestion Control Protocol}
\newacronym{mptcp}{MPTCP}{Multipath TCP}
\newacronym{mr}{MR}{Maximum Rate}
\newacronym{mrdc}{MR-DC}{Multi \gls{rat} \gls{dc}}
\newacronym{mse}{MSE}{Mean Square Error}
\newacronym{mss}{MSS}{Maximum Segment Size}
\newacronym{mt}{MT}{Mobile Termination}
\newacronym{mtd}{MTD}{Machine-Type Device}
\newacronym{mtu}{MTU}{Maximum Transmission Unit}
\newacronym{mumimo}{MU-MIMO}{Multi-user \gls{mimo}}
\newacronym{mvno}{MVNO}{Mobile Virtual Network Operator}
\newacronym{nalu}{NALU}{Network Abstraction Layer Unit}
\newacronym{nas}{NAS}{Non-Access Stratum}
\newacronym{nbiot}{NB-IoT}{Narrow Band IoT}
\newacronym{nfv}{NFV}{Network Function Virtualization}
\newacronym{nfvi}{NFVI}{Network Function Virtualization Infrastructure}
\newacronym{ni}{NI}{Network Interfaces}
\newacronym{nic}{NIC}{Network Interface Card}
\newacronym{nlos}{NLOS}{Non-Line-of-Sight}
\newacronym{now}{NOW}{Non Overlapping Window}
\newacronym{nsm}{NSM}{Network Service Mesh}
\newacronym[type=hidden]{nr}{NR}{New Radio}
\newacronym{nrf}{NRF}{Network Repository Function}
\newacronym{nsa}{NSA}{Non Stand Alone}
\newacronym{nse}{NSE}{Network Slicing Engine}
\newacronym{nssf}{NSSF}{Network Slice Selection Function}
\newacronym{o2i}{O2I}{Outdoor to Indoor}
\newacronym{oai}{OAI}{OpenAirInterface}
\newacronym{oaicn}{OAI-CN}{\gls{oai} \acrlong{cn}}
\newacronym{oairan}{OAI-RAN}{\acrlong{oai} \acrlong{ran}}
\newacronym{oam}{OAM}{Operations, Administration and Maintenance}
\newacronym{ofdm}{OFDM}{Orthogonal Frequency Division Multiplexing}
\newacronym{olia}{OLIA}{Opportunistic Linked Increase Algorithm}
\newacronym{omec}{OMEC}{Open Mobile Evolved Core}
\newacronym{onap}{ONAP}{Open Network Automation Platform}
\newacronym{onf}{ONF}{Open Networking Foundation}
\newacronym{onos}{ONOS}{Open Networking Operating System}
\newacronym{oom}{OOM}{\gls{onap} Operations Manager}
\newacronym{opnfv}{OPNFV}{Open Platform for \gls{nfv}}
\newacronym[type=hidden]{oran}{O-RAN}{O-RAN}
\newacronym{orbit}{ORBIT}{Open-Access Research Testbed for Next-Generation Wireless Networks}
\newacronym{os}{OS}{Operating System}
\newacronym{oss}{OSS}{Operations Support System}
\newacronym{pa}{PA}{Position-aware}
\newacronym{pase}{PASE}{Prioritization, Arbitration, and Self-adjusting Endpoints}
\newacronym{pawr}{PAWR}{Platforms for Advanced Wireless Research}
\newacronym{pbch}{PBCH}{Physical Broadcast Channel}
\newacronym{pcef}{PCEF}{Policy and Charging Enforcement Function}
\newacronym{pcfich}{PCFICH}{Physical Control Format Indicator Channel}
\newacronym{pcrf}{PCRF}{Policy and Charging Rules Function}
\newacronym{pdcch}{PDCCH}{Physical Downlink Control Channel}
\newacronym{pdcp}{PDCP}{Packet Data Convergence Protocol}
\newacronym{pdsch}{PDSCH}{Physical Downlink Shared Channel}
\newacronym{pdu}{PDU}{Packet Data Unit}
\newacronym{pf}{PF}{Proportional Fair}
\newacronym{pgw}{PGW}{Packet Gateway}
\newacronym{phich}{PHICH}{Physical Hybrid ARQ Indicator Channel}
\newacronym{phy}{PHY}{Physical}
\newacronym{pmch}{PMCH}{Physical Multicast Channel}
\newacronym{pmi}{PMI}{Precoding Matrix Indicators}
\newacronym{powder}{POWDER}{Platform for Open Wireless Data-driven Experimental Research}
\newacronym{ppo}{PPO}{Proximal Policy Optimization}
\newacronym{ppp}{PPP}{Poisson Point Process}
\newacronym{prach}{PRACH}{Physical Random Access Channel}
\newacronym{prb}{PRB}{Physical Resource Block}
\newacronym{psnr}{PSNR}{Peak Signal to Noise Ratio}
\newacronym{pss}{PSS}{Primary Synchronization Signal}
\newacronym{pucch}{PUCCH}{Physical Uplink Control Channel}
\newacronym{pusch}{PUSCH}{Physical Uplink Shared Channel}
\newacronym{qam}{QAM}{Quadrature Amplitude Modulation}
\newacronym{qci}{QCI}{\gls{qos} Class Identifier}
\newacronym{qoe}{QoE}{Quality of Experience}
\newacronym{qos}{QoS}{Quality of Service}
\newacronym{quic}{QUIC}{Quick UDP Internet Connections}
\newacronym{rach}{RACH}{Random Access Channel}
\newacronym{ran}{RAN}{Radio Access Network}
\newacronym[firstplural=Radio Access Technologies (RATs)]{rat}{RAT}{Radio Access Technology}
\newacronym{rcn}{RCN}{Research Coordination Network}
\newacronym{rc}{RC}{RAN Control}
\newacronym{rec}{REC}{Radio Edge Cloud}
\newacronym{red}{RED}{Random Early Detection}
\newacronym{renew}{RENEW}{Reconfigurable Eco-system for Next-generation End-to-end Wireless}
\newacronym{rf}{RF}{Radio Frequency}
\newacronym{rfc}{RFC}{Request for Comments}
\newacronym{rfr}{RFR}{Random Forest Regressor}
\newacronym{ric}{RIC}{\gls{ran} Intelligent Controller}
\newacronym{rlc}{RLC}{Radio Link Control}
\newacronym{rlf}{RLF}{Radio Link Failure}
\newacronym{rlnc}{RLNC}{Random Linear Network Coding}
\newacronym{rl}{RL}{Reinforcement Learning}
\newacronym{rmr}{RMR}{RIC Message Router}
\newacronym{rmse}{RMSE}{Root Mean Squared Error}
\newacronym{rnis}{RNIS}{Radio Network Information Service}
\newacronym{rr}{RR}{Round Robin}
\newacronym{rrc}{RRC}{Radio Resource Control}
\newacronym{rrm}{RRM}{Radio Resource Management}
\newacronym{rru}{RRU}{Remote Radio Unit}
\newacronym{rs}{RS}{Remote Server}
\newacronym{rsrp}{RSRP}{Reference Signal Received Power}
\newacronym{rsrq}{RSRQ}{Reference Signal Received Quality}
\newacronym{rss}{RSS}{Received Signal Strength}
\newacronym{rssi}{RSSI}{Received Signal Strength Indicator}
\newacronym{rtt}{RTT}{Round Trip Time}
\newacronym{ru}{RU}{Radio Unit}
\newacronym{rw}{RW}{Receive Window}
\newacronym{rx}{RX}{Receiver}
\newacronym{s1ap}{S1AP}{S1 Application Protocol}
\newacronym{sa}{SA}{standalone}
\newacronym{sack}{SACK}{Selective Acknowledgment}
\newacronym{sap}{SAP}{Service Access Point}
\newacronym{sc2}{SC2}{Spectrum Collaboration Challenge}
\newacronym{scef}{SCEF}{Service Capability Exposure Function}
\newacronym{sch}{SCH}{Secondary Cell Handover}
\newacronym{scoot}{SCOOT}{Split Cycle Offset Optimization Technique}
\newacronym{sctp}{SCTP}{Stream Control Transmission Protocol}
\newacronym{sdap}{SDAP}{Service Data Adaptation Protocol}
\newacronym{sdk}{SDK}{Software Development Kit}
\newacronym{sdm}{SDM}{Space Division Multiplexing}
\newacronym{sdma}{SDMA}{Spatial Division Multiple Access}
\newacronym{sdn}{SDN}{Software-defined Networking}
\newacronym{sdr}{SDR}{Software-defined Radio}
\newacronym{seba}{SEBA}{SDN-Enabled Broadband Access}
\newacronym{sgsn}{SGSN}{Serving GPRS Support Node}
\newacronym{sgw}{SGW}{Service Gateway}
\newacronym{si}{SI}{Study Item}
\newacronym{sib}{SIB}{Secondary Information Block}
\newacronym{sinr}{SINR}{Signal to Interference plus Noise Ratio}
\newacronym{sip}{SIP}{Session Initiation Protocol}
\newacronym{siso}{SISO}{Single Input, Single Output}
\newacronym{sla}{SLA}{Service Level Agreement}
\newacronym{sm}{SM}{Service Model}
\newacronym{smf}{SMF}{Session Management Function}
\newacronym{smo}{SMO}{Service Management and Orchestration}
\newacronym{sms}{SMS}{Short Message Service}
\newacronym{smsgmsc}{SMS-GMSC}{\gls{sms}-Gateway}
\newacronym{snr}{SNR}{Signal-to-Noise-Ratio}
\newacronym{son}{SON}{Self-Organizing Network}
\newacronym{sota}{SOTA}{State-of-the-art}
\newacronym{sptcp}{SPTCP}{Single Path TCP}
\newacronym{srb}{SRB}{Service Radio Bearer}
\newacronym{srn}{SRN}{Standard Radio Node}
\newacronym{srs}{SRS}{Sounding Reference Signal}
\newacronym{ss}{SS}{Synchronization Signal}
\newacronym{sss}{SSS}{Secondary Synchronization Signal}
\newacronym{st}{ST}{Spanning Tree}
\newacronym{svc}{SVC}{Scalable Video Coding}
\newacronym{tb}{TB}{Transport Block}
\newacronym{tbs}{TBS}{Transport Block Size}
\newacronym{tcp}{TCP}{Transmission Control Protocol}
\newacronym{tdd}{TDD}{Time Division Duplexing}
\newacronym{tdm}{TDM}{Time Division Multiplexing}
\newacronym{tdma}{TDMA}{Time Division Multiple Access}
\newacronym{tfl}{TfL}{Transport for London}
\newacronym{tfrc}{TFRC}{TCP-Friendly Rate Control}
\newacronym{tft}{TFT}{Traffic Flow Template}
\newacronym{tgen}{TGEN}{Traffic Generator}
\newacronym{tip}{TIP}{Telecom Infra Project}
\newacronym{tm}{TM}{Transparent Mode}
\newacronym{to}{TO}{Telco Operator}
\newacronym{tr}{TR}{Technical Report}
\newacronym{trp}{TRP}{Transmitter Receiver Pair}
\newacronym{ts}{TS}{Technical Specification}
\newacronym{tti}{TTI}{Transmission Time Interval}
\newacronym{ttt}{TTT}{Time-to-Trigger}
\newacronym{olla}{OLLA}{Outer Loop Link Adaptation}
\newacronym{tx}{TX}{Transmitter}
\newacronym{uas}{UAS}{Unmanned Aerial System}
\newacronym{uav}{UAV}{Unmanned Aerial Vehicle}
\newacronym{udm}{UDM}{Unified Data Management}
\newacronym{udp}{UDP}{User Datagram Protocol}
\newacronym{udr}{UDR}{Unified Data Repository}
\newacronym{ue}{UE}{User Equipment}
\newacronym{uhd}{UHD}{\gls{usrp} Hardware Driver}
\newacronym{ul}{UL}{Uplink}
\newacronym{um}{UM}{Unacknowledged Mode}
\newacronym{uml}{UML}{Unified Modeling Language}
\newacronym{upa}{UPA}{Uniform Planar Array}
\newacronym{upf}{UPF}{User Plane Function}
\newacronym{urllc}{URLLC}{Ultra Reliable and Low Latency Communications}
\newacronym{usa}{U.S.}{United States}
\newacronym{usim}{USIM}{Universal Subscriber Identity Module}
\newacronym{usrp}{USRP}{Universal Software Radio Peripheral}
\newacronym{utc}{UTC}{Urban Traffic Control}
\newacronym{vim}{VIM}{Virtualization Infrastructure Manager}
\newacronym{vm}{VM}{Virtual Machine}
\newacronym{vnf}{VNF}{Virtual Network Function}
\newacronym{volte}{VoLTE}{Voice over \gls{lte}}
\newacronym{voltha}{VOLTHA}{Virtual OLT HArdware Abstraction}
\newacronym{vr}{VR}{Virtual Reality}
\newacronym{vran}{vRAN}{Virtualized \gls{ran}}
\newacronym{vss}{VSS}{Video Streaming Server}
\newacronym{wbf}{WBF}{Wired Bias Function}
\newacronym{wf}{WF}{Waterfilling}
\newacronym{wg}{WG}{Working Group}
\newacronym{wlan}{WLAN}{Wireless Local Area Network}
\newacronym{osm}{OSM}{Open Source \gls{nfv} Management and Orchestration}
\newacronym{pnf}{PNF}{Physical Network Function}
\newacronym{drl}{DRL}{Deep Reinforcement Learning}
\newacronym{mtc}{MTC}{Machine-type Communications}
\newacronym{osc}{OSC}{O-RAN Software Community}
\newacronym{onnx}{ONNX}{Open Neural Network Exchange}
\newacronym{mns}{MnS}{Management Services}
\newacronym{ves}{VES}{\gls{vnf} Event Stream}
\newacronym{ei}{EI}{Enrichment Information}
\newacronym{fh}{FH}{Fronthaul}
\newacronym{fft}{FFT}{Fast Fourier Transform}
\newacronym{laa}{LAA}{Licensed-Assisted Access}
\newacronym{plfs}{PLFS}{Physical Layer Frequency Signals}
\newacronym{ptp}{PTP}{Precision Time Protocol}
\newacronym{cnn}{CNN}{Convolutional Neural Network}
\newacronym{nn}{NN}{Neural Network}
\newacronym{aoa}{AoA}{Angle of Arrival}
\newacronym{xr}{XR}{Extended Reality}
\newacronym{icc}{ICC}{Intelligence Coordination Controller}
\newacronym{smos}{SMOS}{\gls{smo} Services}
\newacronym{focom}{Federated O-Cloud Orchestration and Management}{FOCOM}
\newacronym{fec}{FEC}{Forward Error Correcting Code}
\newacronym{nfo}{Network Function Orchestration}{NFO}
\newacronym{dms}{DMS}{Deployment Management Service}
\newacronym{llm}{LLM}{Large Language Model}
\newacronym{la}{LA}{Link Adaptation}
\newacronym{ota}{OTA}{Over-the-Air}
\newacronym{roc}{ROC}{Receiver Operating Characteristic}
\newacronym{tdl}{TDL}{Tapped Delay Line}
\newacronym{cdl}{CDL}{Clustered Delay Line}

%% file: references.bib
@IEEEtranBSTCTL{IEEEexample:BSTcontrol,
CTLuse_forced_etal       = "yes",
CTLmax_names_forced_etal = "3",
CTLnames_show_etal       = "2" }

@misc{maxenti2025autoranautomatedzerotouchopen,
      title={{AutoRAN: Automated and Zero-Touch Open RAN Systems}}, 
      author={Stefano Maxenti and Ravis Shirkhani and Maxime Elkael and Leonardo Bonati and Salvatore D'Oro and Tommaso Melodia and Michele Polese},
      year={2025},
      eprint={2504.11233},
      archivePrefix={arXiv},
      primaryClass={cs.NI},
      url={https://arxiv.org/abs/2504.11233}, 
}

@misc{mgen,
author="{U.S. Naval Research Laboratory}",
title="{MGEN Traffic Emulator}",
url_={https://tinyurl.com/beexe8yc},
url={https://www.nrl.navy.mil/Our-Work/Areas-of-Research/Information-Technology/NCS/MGEN},
year={Accessed 2024}
}

@ARTICLE{villa2024x5g,
  author  = {Villa, Davide and Khan, Imran and Kaltenberger, Florian and Hedberg, Nicholas and Soares da Silva, R{\'u}ben and Maxenti, Stefano and Bonati, Leonardo and Kelkar, Anupa and Dick, Chris and Baena, Eduardo and Jornet, Josep M. and Melodia, Tommaso and Polese, Michele and Koutsonikolas, Dimitrios},
  title   = {{X5G}: An Open, Programmable, Multi-Vendor, End-to-End, Private {5G} {O-RAN} Testbed With {NVIDIA ARC} and {OpenAirInterface}},
  journal = {IEEE Transactions on Mobile Computing},
  volume  = {24},
  number  = {11},
  pages   = {11305--11322},
  month   = {November},
  year    = {2025},
  doi     = {10.1109/TMC.2025.3580764}
}

@ARTICLE{lacava2025dapps,
author={Andrea Lacava and Leonardo Bonati and Niloofar Mohamadi and Rajeev Gangula and Florian Kaltenberger and Pedram Johari and Salvatore D'Oro and Francesca Cuomo and Michele Polese and Tommaso Melodia},
title= {{dApps: Enabling Real-Time AI-Based Open RAN Control}},
journal={Computer Networks},
volume={269},
pages={111342},
year={2025},
doi={10.1016/j.comnet.2025.111342}
}

@techreport{3gpp_901,
  author      = {{3GPP}},
  title       = {{TR} 38.901 {5G}; {Study on channel model for frequencies from 0.5 to 100 {GHz}} (3GPP TR 38.901 version 19.4.0 Release 19)},
  institution = {{3rd Generation Partnership Project}},
  year        = {2026},
  month       = {July},
  url         = {https://www.etsi.org}
}

@INPROCEEDINGS{pedersen2007frequency,
  author    = {Pedersen, Klaus I. and Monghal, Guillaume and Kovacs, Istvan Z. and Kolding, Troels E. and Pokhariyal, Akhilesh and Frederiksen, Frank and Mogensen, Preben},
  title     = {Frequency Domain Scheduling for {OFDMA} with Limited and Noisy Channel Feedback},
  booktitle = {Proc. IEEE 66th Vehicular Technology Conference (VTC-2007 Fall)},
  pages     = {1792--1796},
  year      = {2007},
  doi       = {10.1109/VETECF.2007.378}
}

@ARTICLE{blanquez2016eolla,
  author  = {Bl{\'a}nquez-Casado, Francisco and G{\'o}mez, Gerardo and Aguayo-Torres, Mari Carmen and Entrambasaguas, Jos{\'e} T.},
  title   = {{eOLLA}: an enhanced outer loop link adaptation for cellular networks},
  journal = {EURASIP Journal on Wireless Communications and Networking},
  volume  = {2016},
  number  = {1},
  pages   = {20},
  year    = {2016},
  doi     = {10.1186/s13638-016-0518-3}
}

@ARTICLE{wiesmayr2025salad,
  author  = {Wiesmayr, Reinhard and Maggi, Lorenzo and Cammerer, Sebastian and Hoydis, Jakob and A{\"i}t Aoudia, Fay{\c{c}}al and Keller, Alexander},
  title   = {{SALAD}: Self-Adaptive Link Adaptation},
  journal = {arXiv preprint arXiv:2510.05784},
  year    = {2025}
}

@ARTICLE{wang2026lolla,
  author  = {Wang, Rui and Zhang, Linchao and Liu, Qiang and Yang, Kun},
  title   = {{LOLLA}: Deep Reinforcement Learning for Closed-Loop Link Adaptation Towards a {GPU}-Accelerated {AI-RAN}},
  journal = {arXiv preprint arXiv:2606.23110},
  year    = {2026}
}

@ARTICLE{you2026dcdqn,
  author  = {You, Lizhao and Zhou, Nanqing and Pang, Guanglong and Huang, Jiajie and Shao, Yulin and Fu, Liqun},
  title   = {From Simulation to Reality: Practical Deep Reinforcement Learning-based Link Adaptation for Cellular Networks},
  journal = {arXiv preprint arXiv:2603.00689},
  year    = {2026}
}

@ARTICLE{yin2020cqi,
  author  = {Yin, Hao and Guo, Xiaojun and Liu, Pengyu and Hei, Xiaojun and Gao, Yayu},
  title   = {Predicting Channel Quality Indicators for {5G} Downlink Scheduling in a Deep Learning Approach},
  journal = {arXiv preprint arXiv:2008.01000},
  year    = {2020},
  doi     = {10.48550/arXiv.2008.01000}
}

@ARTICLE{tsipi2024mcs,
  author  = {Tsipi, Lefteris and Karavolos, Michail and Papaioannou, Grigorios and Volakaki, Maria and Vouyioukas, Demosthenes},
  title   = {Machine learning-based methods for {MCS} prediction in {5G} networks},
  journal = {Telecommunication Systems},
  volume  = {86},
  number  = {4},
  pages   = {705--728},
  year    = {2024},
  doi     = {10.1007/s11235-024-01158-x}
}

@INPROCEEDINGS{zhu2023nolla,
  author    = {Zhu, Lingrui and Bockelmann, Carsten and Schier, Thorsten and Hajri, Salah Eddine and Dekorsy, Armin},
  title     = {{NOLLA}: Non-Linear Outer Loop Link Adaptation for Enhancing Wireless Link Transmission},
  booktitle = {2023 IEEE 34th Annual International Symposium on Personal, Indoor and Mobile Radio Communications (PIMRC)},
  pages     = {1--6},
  year      = {2023},
  doi       = {10.1109/PIMRC56721.2023.10293829}
}

@ARTICLE{saxena2022rlla,
  author  = {Saxena, Vidit and Tullberg, Hugo M. and Jald{\'e}n, Joakim},
  title   = {Reinforcement Learning for Efficient and Tuning-Free Link Adaptation},
  journal = {IEEE Transactions on Wireless Communications},
  volume  = {21},
  number  = {2},
  pages   = {768--780},
  year    = {2022},
  doi     = {10.1109/TWC.2021.3098972}
}

@ARTICLE{ye2023drlla,
  author  = {Ye, Xiaowen and Yu, Yiding and Fu, Liqun},
  title   = {Deep Reinforcement Learning Based Link Adaptation Technique for {LTE/NR} Systems},
  journal = {IEEE Transactions on Vehicular Technology},
  volume  = {72},
  number  = {6},
  pages   = {7364--7379},
  year    = {2023},
  doi     = {10.1109/TVT.2023.3236791}
}

@ARTICLE{friedman2001greedy,
  author  = {Friedman, Jerome H.},
  title   = {Greedy Function Approximation: A Gradient Boosting Machine},
  journal = {The Annals of Statistics},
  volume  = {29},
  number  = {5},
  pages   = {1189--1232},
  year    = {2001}
}

@ARTICLE{diazruiz2025csi,
  author  = {D{\'i}az-Ruiz, Francisco and Mart{\'i}n-Vega, Francisco J. and Cort{\'e}s, Jos{\'e} A. and G{\'o}mez, Gerardo and Aguayo, Mari Carmen},
  title   = {{CSI} Prediction Frameworks for Enhanced {5G} Link Adaptation: Performance-Complexity Trade-offs},
  journal = {arXiv preprint arXiv:2511.20160},
  year    = {2025},
  doi     = {10.48550/arXiv.2511.20160}
}

@INPROCEEDINGS{deshpande2024openairlink,
  author    = {Deshpande, Yash and Wang, Xianglong and Kellerer, Wolfgang},
  title     = {{OpenAirLink}: Reproducible Wireless Channel Emulation using Software Defined Radios},
  booktitle = {Proc. IFIP Networking Conference},
  pages     = {678--683},
  year      = {2024},
  doi       = {10.23919/IFIPNetworking62109.2024.10619070}
}
